\documentclass[fleqn,usenatbib]{mnras}

\usepackage{newtxtext,newtxmath}

\usepackage[T1]{fontenc}
\usepackage{ae,aecompl}

\usepackage{graphicx}	
\usepackage{amsmath}	
\usepackage{amssymb}	

\title[Photon-induced processes on a pure CH$_4$ ice]{Photon-induced desorption of larger species in UV-irradiated methane (CH$_4$) ice}

\author[H. Carrascosa et al.]{
H. Carrascosa,$^{1}$\thanks{E-mail: hcarrascosa@cab.inta-csic.es}
G.A. Cruz-D\'iaz,$^{2,3}$
G. M. Mu\~noz Caro,$^{1}$\thanks{E-mail: munozcg@cab.inta-csic.es}
E. Dartois,$^{4}$
Y. -J. Chen$^{5}$\thanks{E-mail: asperchen@phy.ncu.edu.tw}
\\
$^{1}$Centro de Astrobiolog\'{\i}a (CSIC-INTA), Ctra. de Ajalvir, km 4, Torrej\'on de Ardoz, 28850 Madrid, Spain\\
$^{2}$NASA Ames Research Center, Moffett Field, Mountain View, CA 94035, USA\\
$^{3}$Bay Area Environmental Research Institute, Moffett Field, Mountain View, CA 94035, USA\\
$^{4}$Institut des Sciences Mol\`eculaires d'Orsay (ISMO), UMR8214, CNRS - Universit\'e de Paris-Sud, Universit\'e Paris-Saclay, B\^at 520, Rue Andr\'e Rivi\`ere,\\
91405 Orsay, France\\
$^{5}$Department of Physics, National Central University, Jhongli City, Taoyuan County 32054, Taiwan\\
}

\date{Accepted XXX. Received YYY; in original form ZZZ}

\pubyear{2019}

\hypersetup{draft}
\begin{document}
\label{firstpage}
\pagerange{\pageref{firstpage}--\pageref{lastpage}}
\maketitle

\begin{abstract}
At the low temperatures found in the interior of dense clouds and circumstellar regions, along with H$_2$O and smaller amounts of species such as CO, CO$_2$, or CH$_3$OH, the infrared features of CH$_4$ have been observed on icy dust grains. Ultraviolet (UV) photons induce different processes in ice mantles, affecting the molecular abundances detected in the gas-phase.\\
This work aims to understand the processes that occur in a pure CH$_4$ ice mantle submitted to UV irradiation. We studied photon-induced processes for the different photoproducts arising in the ice upon UV irradiation.\\
Experiments were carried out in ISAC, an ultra-high vacuum chamber equipped with a cryostat and an F-type UV-lamp reproducing the secondary UV-field induced by cosmic rays in dense clouds. Infrared spectroscopy and quadrupole mass spectrometry were used to monitor the solid and gas-phase, respectively, during the formation, irradiation, and warm-up of the ice.\\
Direct photodesorption of pure CH$_4$ was not observed. UV photons form CH$_x\cdot$ and H$\cdot$ radicals, leading to photoproducts such as H$_2$, C$_2$H$_2$, C$_2$H$_6$, and C$_3$H$_8$. Evidence for the  photodesorption of C$_2$H$_2$ and photochemidesorption of C$_2$H$_6$ and C$_3$H$_8$ was found, the latter species is so far the largest molecule found to photochemidesorb. $^{13}$CH$_4$ experiments were also carried out to confirm the reliability of these results.\\
\end{abstract}




\section{Introduction}
\label{Introduction}
The low temperatures present in dense interstellar clouds promote the formation of ice mantles around carbonaceous and silicate dust grains. In the outer regions of the cloud, the temperature and the UV field are too high for any molecule to freeze out. Deeper inside the cloud, there is sufficient screening from the external UV radiation field, and molecules form and accrete onto dust grains, thus forming ice mantles. Even the highly volatile species like CO and N$_2$ are frozen at 10 K. Methane has been detected in the interstellar medium (ISM) and dense clouds \citep{Lacy1991,Boogert1996,Oberg2008}. Methane ice can be formed from successive hydrogenation of carbon atoms over a dust grain surface, from photoprocessing of CH$_3$OH ice, or even from gas-phase reactions and subsequent freeze out over dust grains \citep[and references therein]{Oberg2008}. CH$_4$ constitutes a source of carbon atoms in ice mantles, with abundances around 5\% and 2\% of the water ice in low-mass and high-mass protostars, respectively \citep{Dartois2005SSRv..119..293D, Oberg2011, Boogert2015ARA&A..53..541B}. Complex organic molecules (COMs), containing six or more atoms and at least one carbon, can be formed from methane processing in the interstellar and circumstellar medium, or comets. These systems contain variable quantities, up to 4\% relative to water, of solid methane, which is exposed to vacuum ultraviolet (UV) photons with a spectral energy distribution as simulated in our experiments \citep{Gerakines1996, Lin2014}. In our Solar System, CH$_4$ has been detected on the surface of Triton \citep{Cruikshank1993, Owen1993}, Titan \citep{McKay1997}, and Pluto \citep{Grundy2016}, among others.\\

At the low temperatures that govern the interior of dense clouds, thermal energy is almost negligible, and ice processes driven by cosmic rays and UV photons play a significant role. Secondary UV radiation, generated by the interaction between cosmic rays and hydrogen molecules \citep{Cecchi1992, Shen2004} is the main source of photons impinging on ice mantles in dense clouds. The absorption of UV photons by methane molecules in the ice leads to an efficient photochemistry. In this article, we explore the photon-induced desorption of molecules upon UV-irradiation of CH$_4$ in the laboratory.\\

The absorption of sufficient photon energy by a molecule on the ice surface counteracts the intermolecular forces with the surrounding species, allowing photon-induced desorption. UV photons, however, can also break covalent bonds, inducing the formation of radicals that may react to form new molecules, leading to photochemistry. Depending on the species present in the ice and the photon energy, UV radiation induces photon-induced desorption, photochemistry, or both.\\

The direct desorption of a molecule from the ice surface \citep{vanHemert2015, Dupuy2017}, after absorption of a photon, is of low efficiency compared to the indirect photodesorption, by which the photon is absorbed by another molecule and the photon energy is distributed to the surrounding molecules. In this case, a UV photon can be absorbed by a molecule in the subsurface layers of the ice, leading to electronic excitation. The energy from the relaxation to the ground state is redistributed to the neighboring molecules. A different kind of photon-induced desorption was observed when a surface molecule absorbs a photon and dissociates into photofragments that may recombine, forming a photoproduct. This photoproduct can desorb if the  excess energy of the parent photofragments is sufficient to overcome the binding energy of the ice, due to the exothermicity of the reaction \citep{Andersson2008, Fayolle2013, Fillion2014, Bertin2016}. Photochemical desorption or photochemidesorption only applies to species that are formed on the ice surface and are immediately ejected to the gas, leading to a constant photodesorption yield during the irradiation \citep{MD2016, MD2018, Gus2016}.\\

The IR features arising from UV radiation or $\alpha$-particle and proton bombardment of pure CH$_4$ ice were reported \citep{Gerakines1996, Kaiser1998}. More recently, \cite{Lin2014} and \cite{Lo2015MNRAS.451..159L} irradiated CH$_4$ ice at 3 K with monochromatic UV light from a synchrotron, and \cite{Dupuy2017} studied CH$_4$ photodesorption using monochromatic UV radiation. This work focuses on photon-induced desorption processes of CH$_4$ photoproducts using a continuum UV emission lamp that mimics the secondary UV-field in dense clouds. We will show that photoproducts desorb mainly via photochemidesorption.\\

\section{Experimental}
\label{experimental_setup}
Experiments were carried out in the InterStellar Astrochemistry Chamber (ISAC) located at the Centro de Astrobiolog\'ia (for a detailed description of ISAC see \cite{Guille2010}). ISAC is an ultra-high vacuum chamber with a base pressure around 4$\times$10$^{-11}$ mbar, similar to the pressure in dense interstellar clouds. A closed-cycle helium cryostat allows to cool down to 8 K, a temperature similar to that of dust grains in the inner parts of dense interstellar clouds. CH$_4$ (gas, Praxair 99.95\%) and $^{13}$CH$_4$ (gas, Cambridge Isotopes Laboratories 99.9\%) were used for the experiments. Gases at a pressure of 2$\times$10$^{-7}$ mbar were introduced in the UHV-chamber through a stainless steel tube at normal incidence angle with respect to the MgF$_2$ substrate. The low temperature of the substrate led to the formation of amorphous ices \citep{Hudson2015}. The column density of the ice samples corresponds to an absorption of about 90\% of the incident photon flux in the deposited ice layers. To estimate the number of UV photons absorbed, we adopted the UV absorption cross-section of CH$_4$ ice measured by \cite{Gus2014b}.\\

Fourier-Transform Infrared Spectroscopy (FTIR) measured with a Bruker Vertex 70 spectrometer equipped with a deuterated triglycine sulfate detector (DTGS) was used to monitor the column densities and composition of the ice samples. IR spectra were measured with a resolution of 2 cm$^{-1}$ after deposition of the ice, after each irradiation step, and also during the warm-up. The measured spectral range in the reported experiments spans from 6000 to 1200 cm$^{-1}$ (1.66 to 8.33 $\mu$m), since the MgF$_2$ substrate becomes optically thick at longer wavelengths. The column density of the ice layer was determined using equation \ref{Eq.Band.strength}, where $N$ is the column density of the molecule of interest in molec $\cdot$ cm$^{-2}$, $A$ is the band strength of the considered band in cm $\cdot$ molecule$^{-1}$, $\tau _{v}$ the optical depth of the band, and $dv$ the wavenumber differential in cm$^{-1}$:\\

\begin{equation}
\centering
\hspace*{8em}N = \frac{1}{A} \int_{band}{\tau _{v}\; dv.}
\label{Eq.Band.strength}
\end{equation}\\

Ice samples were irradiated using an F-type microwave discharge hydrogen lamp (MDHL) from Opthos Instruments, providing an UV flux of 2$\times$10$^{14}$ photons $\cdot$ cm$^{-2}$ $\cdot$ s$^{-1}$ and 8.6 eV average photon energy (e. g. \cite{Gus2016}). The MgF$_2$ window between the MDHL and the vacuum chamber leads to a cut-off at 114 nm (10.87 eV). Destruction cross-section of CH$_4$, and similarly of $^{13}$CH$_4$, were obtained using equation \ref{Eq.Destruction_Cross_Section}, where $N_t(\rm{CH_4})$ and $N_0(\rm{CH_4})$ are the column densities measured by FTIR before and after the irradiation, respectively, in molec $\cdot$ cm$^{-2}$, $\phi$ is the UV flux in photons $\cdot$ cm$^{-2}$ $\cdot$ s$^{-1}$, $t$ is the irradiation time in s, which belongs to the first irradiation period, as the ice is less processed at the beginning, and $\sigma_{des}(\rm{CH_4})$ is the destruction cross-section in cm$^{2}$.\\

\begin{equation}
\centering
\hspace*{6em} N_t(\rm{CH_4}) = N_0(\rm{CH_4}) \times e^{- \phi t \sigma_{des}(\rm{CH_4})}
\label{Eq.Destruction_Cross_Section}
\end{equation}\\

Formation cross-section for ethane molecules ($\sigma_{for}(\rm{C_2H_6})$) was obtained using equation \ref{Eq.Formation_Cross_Section}, where $\frac{dN(\rm{C_2H_6})}{dt}$ is the variation of the column density of ethane molecules after irradiation time, $t$ is the irradiation time in s, and $N(\rm{CH_4})$ is the average column density of methane during the irradiation interval.\\

\begin{equation}
\centering
\hspace*{4em}\frac{dN(\rm{C_2H_6})}{dt} = - \phi \times t \times N(\rm{CH_4}) \times \sigma_{for}(\rm{C_2H_6})
\label{Eq.Formation_Cross_Section}
\end{equation}\\

The gas-phase was monitored during deposition, irradiation, and warm-up processes using a Pfeiffer Prisma quadrupole mass spectrometer (QMS) equipped with a Channeltron detector. Molecules were ionized by electron impact of 70 eV and accelerated into the mass filter of the QMS, leading to a characteristic fragmentation pattern for each species. From the ion current obtained in the QMS for each molecular fragment, a quantification of the gas-phase molecules was done following the procedure presented in \cite{MD2015}. After baseline and blank (no ice sample) irradiation corrections, the area below the QMS during each irradiation period was calculated, obtaining the ion current as a function of the photon dose. Applying Eq. \ref{Eq.QMS} and the values in Table \ref{Table.QMS}, photodesorption rates of methane, ethane and propane were obtained by monitoring $\frac{m}{z}=$ 15, $\frac{m}{z}=$ 30, and $\frac{m}{z}=$ 43, respectively. These $\frac{m}{z}$ fragments were selected considering the possible contribution of other molecules and the relative intensity for each of the fragments. In Eq. \ref{Eq.QMS}, $N(mol)$ is the photodesorption rate in molec $\cdot$ photon$^{-1}$, $A\left(\frac{m}{z}\right)$ is the area obtained from the QMS normalized by the number of incident photons, $k_{CO}$ is the proportionality constant from the calibration of the QMS in a CO ice irradiation experiment (see \cite{MD2015}), $\sigma^{+}$(mol) is the ionization cross-section of each species ionized at a voltage of 70 eV in the QMS (data adopted from the National Institute of Standards and Technology), $IF(z)$ is the ionization factor, which has been considered unity for all molecules, as most of them will be charged once, $FF$ is the fragmentation factor, it refers to the percentage of molecules which account for a given $\left(\frac{m}{z}\right)$ value (using the mass spectrum of methane measured by our QMS during deposition, while the mass spectra from NIST database was used for the photoproducts), and $S\left(\frac{m}{z}\right)$ is the sensitivity of the QMS, which depends on the ion current generated at the QMS as a function of the mass of each molecule (for calculation details, see \cite{MD2015}).\\

\begin{table} 
    \caption[]{Values adopted for the calculation of photodesorption rates by QMS}
    \label{Table.QMS}
    \begin{center}
    \begin{tabular}{ccccc}
Parameter   & CO & CH$_4$ & C$_2$H$_6$ & C$_3$H$_8$\\
\noalign{\smallskip}
\hline
\hline
\noalign{\smallskip}
\noalign{\smallskip}
$k_{CO}$ $\left(\frac{A \cdot min}{molecule}\right)$ & $1.32\cdot10^{5}$ &-&-&-\\
\noalign{\smallskip}
\noalign{\smallskip}
$\sigma^{+}$(mol) (A$^{2}$) & 2.516 & 3.524 & 6.422 & 8.619\\
\noalign{\smallskip}
$FF$ & 0.939 & 0.419 & 0.122* &0.0757*\\
\noalign{\smallskip}
\noalign{\smallskip}
\hline
\end{tabular}\\
\end{center}
\textit{* Values taken from the National Institute of Standards and Technology (NIST) database.}
\end{table}

\begin{equation}
  \hspace*{0.7em}N(mol) = \frac{A\left(\frac{m}{z}\right)}{k_{CO}} \cdot \frac{\sigma^{+}(CO)}{\sigma^{+}(mol)} \cdot
        \frac{IF(CO^{+})}{IF(z)} \cdot \frac{FF(28)}{FF(m)} \cdot \frac{S(28)}{S(\frac{m}{z})} \\
  \label{Eq.QMS}
\end{equation}\\
After the irradiation steps, ice samples were warmed up until room temperature using a LakeShore Model 331 temperature controller. The temperature was monitored by a silicon diode sensor with a sensitivity of 0.1 K, located underneath the sample holder.\\

\section{Results and discussion}
\label{Results}

\setcounter{equation}{0} 
\renewcommand{\theequation}{\roman{equation}} 
\subsection{Photochemical processes}
\label{FTIR}

\begin{figure} 
  \centering 
  \includegraphics[width=0.5\textwidth]{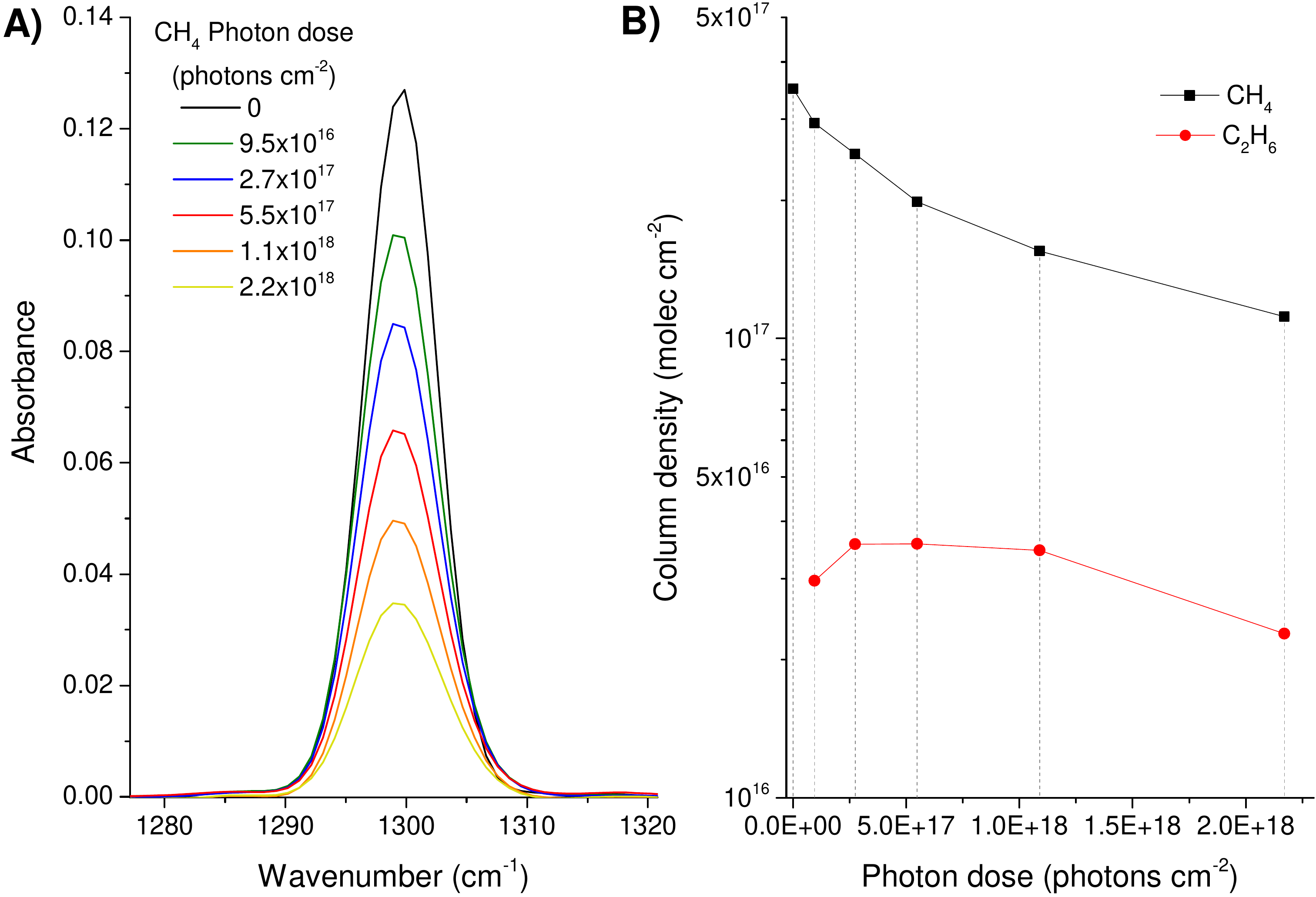}
  \caption{A) Evolution of the $\nu_3$ vibration mode of methane as a function of the photon dose for CH$_4$. B) Column density calculated for CH$_4$ and C$_2$H$_6$ after each irradiation period.}
  \label{Fig.IR}
\end{figure}

Fig. \ref{Fig.IR}A shows the evolution of the $\nu_3$ stretching mode of CH$_4$ (1301 cm$^{-1}$). The reaction network presented in scheme \ref{Sch.CH4} shows that UV photons can dissociate methane molecules producing CH$_3\cdot$ and H$\cdot$ radicals, CH$_2\cdot$ radicals and H$_2$ molecules or, to a lesser extent, CH$\cdot$ radicals, H$_2$ and H$\cdot$ radicals. The position of the $\nu_3$ IR band does not change, while its intensity is reduced. In the gas-phase, for 123.6 nm (10.03 eV) photons, the quantum yields for the second and third reactions are 0.5 and 0.06, respectively \citep{Okabe1978}.\\

\begin{eqnarray}
\begin{split}
\label{Sch.CH4}
\centering
  &\rm{CH}_4 \thinspace + \thinspace h\nu \medspace \xrightarrow{} \thinspace \rm{CH}_3 \cdot \medspace + \;\;\medspace \rm{H} \cdot\\
  &\rm{CH}_4 \thinspace + \thinspace h\nu \medspace \xrightarrow{} \thinspace \rm{CH}_2 \cdot \medspace + \;\;\medspace \rm{H}_2\\
  &\rm{CH}_4 \thinspace + \thinspace h\nu \medspace \xrightarrow{} \thinspace \rm{CH} \cdot \medspace + \;\;\medspace \rm{H}_2 \medspace + \;\;\medspace \rm{H} \cdot\\ 
\end{split}
\end{eqnarray}\\

Radicals produced from methane dissociation will recombine giving rise to new species. This observation is supported by the formation of hydrogenated amorphous carbon (a-C:H) at 10 K in similar CH$_4$ irradiation experiments \citep{Dartois2005}. As a consequence, bands related to CH$_x\cdot$ radicals were not detected in IR spectra (e. g. the 3150 cm$^{-1}$ band for CH$_3\cdot$, \cite{deBarros2011}). H$\cdot$ radicals can diffuse inside the ice and recombine producing H$_2$ molecules, as shown in Scheme \ref{Sch.H2}.\\

\begin{eqnarray}
\label{Sch.H2}
   \rm{H} \cdot \thinspace + \medspace \rm{H} \cdot \medspace \xrightarrow{} \thinspace \rm{H}_2  
\end{eqnarray}\\

\begin{figure} 
  \centering 
  \includegraphics[width=0.5\textwidth]{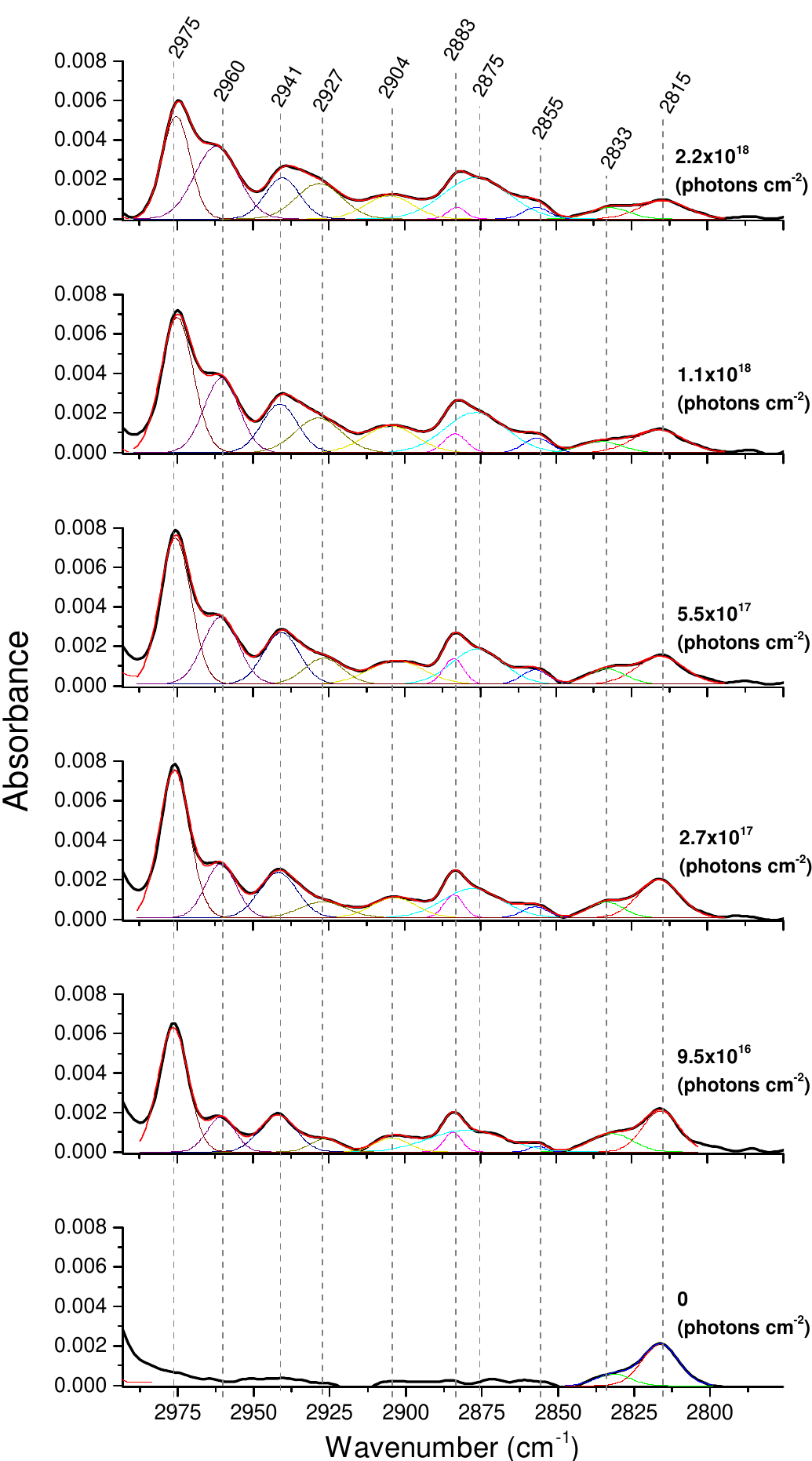}
  \caption{Evolution of CH$_4$ IR features between 3000-2800 cm$^{-1}$ under UV irradiation. Black trace belongs to the original spectra. Red trace represents the sum of 10 Gaussian profiles. Gaussian profiles were used to take peak positions and calculate the abundances.}
  \label{Fig.Gaussian}
\end{figure}

The reaction of two CH$_3\cdot$ radicals in the gas-phase produces ethane, C$_2$H$_6$, (Scheme \ref{Sch.C2H6}), with a rate constant of 5 $\times$ 10$^{-11}$ cm$^{3}$ molec$^{-1}$ s$^{-1}$ \citep{Okabe1978}, leading to the appearance of its characteristic IR bands (see Fig. \ref{Fig.Gaussian} and Table. \ref{Table.1}). Ethane can also be produced from the reaction between a CH$_2\cdot$ radical and a CH$_4$ molecule, although the lower formation rate of CH$_2\cdot$ radicals would probably determine a minor production rate by this route.

\begin{eqnarray}
\begin{split}
\label{Sch.C2H6}
  &\rm{CH}_3 \cdot \thinspace + \medspace \rm{CH}_3 \cdot \medspace \xrightarrow{} \thinspace \rm{C}_2\rm{H}_6\\
  &\rm{CH}_2 \cdot \thinspace + \medspace \rm{CH}_4 \medspace \xrightarrow{} \thinspace \rm{C}_2\rm{H}_6
\end{split}
\end{eqnarray}\\

As shown in Fig. \ref{Fig.Gaussian}, bands related to C-H stretching modes from different molecules appeared in the IR spectrum between 3000-2800 cm$^{-1}$ upon irradiation. The profile of this region was fitted using 10 Gaussian curves. The bands assignment is shown in Table \ref{Table.1} for CH$_4$ and $^{13}$CH$_4$ irradiations. Within the first steps of irradiation, the integrated IR absorption of the photoproducts increases as a consequence of the ice processing. However, IR bands related to ethane (2975, 2941 and 2883 cm$^{-1}$) decrease for longer irradiation times, while other IR bands are still growing. The formation of hydrogen molecules, which can escape from the ice, determines that the proportion of carbon to hydrogen increases during the irradiation, and CH$_3\cdot$ radicals are more and more scarce. Therefore,  the production of C$_2$H$_6$ (Scheme \ref{Sch.C2H6}) molecules is lowered, whereas its dissociation (Scheme \ref{Sch.C2H6.2}) does not change, in agreement with the observed ethane band intensities.

\begin{table*} 
    \caption[]{Main IR features detected between 3100-2800 cm$^{-1}$ after 240 min irradiation of a pure CH$_4$ ice (left) and a pure $^{13}$CH$_4$ ice (right). Positions belong to the centre of the Gaussian profiles, shown in Fig. \ref{Fig.Gaussian} for a CH$_4$ experiment. $^{13}$CH$_4$ IR spectra (not shown) revealed a similar profile.}
    \label{Table.1}
    \begin{center}
    \begin{tabular}{cccccccc}
\multicolumn{3}{c}{CH$_4$}           && \multicolumn{3}{c}{$^{13}$CH$_4$}\\
\cline{1-3}
\cline{5-7}
\noalign{\smallskip}
\noalign{\smallskip}
Frequency (cm$^{-1}$)  & Asignment & Reference &&Frequency (cm$^{-1}$) &Asignment &Reference\\
\noalign{\smallskip}
\noalign{\smallskip}
\cline{1-3}
\cline{5-7}
&&&\;\;\;\;\;\;\;\\
\noalign{\smallskip}
\noalign{\smallskip}
3009        &CH$_4$         &1, 6       &&3000       &$^{13}$CH$_4$     &7\\
\noalign{\smallskip}
2975        &C$_2$H$_6$     &1, 2, 8       &&2966       &$^{13}$C$_2$H$_6$ &7\\
\noalign{\smallskip}
2960        &R-CH$_2$-R'     &2, 3, 9       &&2958       &R-$^{13}$CH$_3$   &7\\
\noalign{\smallskip}
2941        &C$_2$H$_6$     &2          &&2940       &$^{13}$C$_2$H$_6$ &This work\\
\noalign{\smallskip}
2927        &R-CH$_2$-R'    &4          &&2931       &R-$^{13}$CH$_2$-R     &7\\
\noalign{\smallskip}
2904        &?              &           &&2909       &?\\
\noalign{\smallskip}
2883        &C$_2$H$_6$	    &1, 2, 3, 5, 8    &&2879    &R,R'-$^{13}$CH-R'' / $^{13}$C$_2$H$_6$            &7\\
\noalign{\smallskip}
2875        &R-CH$_3$       &5          &&2871    &R-$^{13}$CH$_3$      &This work\\
\noalign{\smallskip}
2855        &R-CH$_2$-R'    &2          &&2843    &R-$^{13}$CH$_2$-R'    &This work\\
\noalign{\smallskip}
2833        &R-CH$_3$       &5          &&2817    &R-$^{13}$CH$_3$      &This work\\
\noalign{\smallskip}
2815        &CH$_4$         &1, 7       &&2807    &$^{13}$CH$_4$        &This work\\
\noalign{\smallskip}
\noalign{\smallskip}
\hline
\noalign{\smallskip}
\end{tabular}\\
\end{center}
\textit{
1- \cite{Gerakines1996}; 2- \cite{deBarros2011}; 3- \cite{Moore2003}; 4- \cite{Dartois2005}; 5- \cite{Bennett2006}; 6- \cite{d'Hendecourt1986A&A...158..119D}; 7- \cite{Kaiser1998}; 8- \cite{Lin2014}; 9- \cite{Baratta2002}}
\end{table*}

\begin{eqnarray}
\label{Sch.C2H6.2}
  &\rm{C}_2\rm{H}_6 \thinspace + \thinspace h\nu \medspace \xrightarrow{} \thinspace \rm{C}_2\rm{H}_5 \cdot \medspace + \;\;\medspace \rm{H} \cdot
\end{eqnarray}\\

The CH$_4$ band at 1301 cm$^{-1}$ was selected for quantification of its column density using Eq. \ref{Eq.Band.strength} and assuming a band strength of 6.1x10$^{-18}$ cm $\cdot$ molecule$^{-1}$ \citep{Gerakines1996, Kaiser1998}. Fig. \ref{Fig.IR}B shows the column density of methane and ethane as a function of the incident photon flux. The Gaussian profile at 2975 cm$^{-1}$ was used to obtain the column density of C$_2$H$_6$ molecules, using a band strength of 1.3x10$^{17}$ cm $\cdot$ molecule$^{-1}$, following \cite{Gerakines1996}. Taking into account the destruction cross-section of methane shown in Table \ref{Table.sigmas}, and the column density of ethane molecules after the first irradiation step, the formation cross-sections of C$_2$H$_6$ was obtained using equation \ref{Eq.Formation_Cross_Section} (see Table \ref{Table.sigmas}). Ethane was found to be the main molecular photoproduct for short irradiation periods, but, as the amount of CH$_3\cdot$ radicals is reduced, less hydrogenated compounds arise in the ice upon further irradiation.\\

\begin{table} 
    \caption[]{Column density and destruction/formation cross-section ($\sigma$) of methane and ethane for CH$_4$ ice samples. A 10\% error was estimated by the sum of the errors from the column density, and the MDHL, photomultiplier and ammeter stabilities for the incident cross-sections, whereas in the absorption cross-section presented in \cite{Gus2014b}, a 20\% error is assumed for the absorbed cross-sections.}
    \label{Table.sigmas}
    \begin{center}
    \begin{tabular}{cccc}
N$_{t=0}$ (molec$\cdot$cm$^{-2}$) & photon flux ($\phi$) & Species &  $\sigma$ (cm$^{2}$)\\
\hline
\noalign{\smallskip}
3.5$\times$10$^{17}$ & incident &CH$_4$ &1.9 x10$^{-18}$\\
\noalign{\smallskip}
&&C$_2$H$_6$  &1.0 x10$^{-18}$\\
\noalign{\smallskip}
3.5$\times$10$^{17}$ & absorbed &CH$_4$ &2.2 x10$^{-18}$\\
\noalign{\smallskip}
&&C$_2$H$_6$  &1.2 x10$^{-18}$\\
\noalign{\smallskip}
\hline
\end{tabular}\\
\end{center}
\end{table}


Dissociation of ethane molecules gives rise to the formation of other photoproducts. C$_2$H$_5\cdot$ radicals produced from ethane dissociation may react with CH$_3\cdot$ radicals, thus forming propane (Scheme \ref{Sch.C3H8}). Propane can also be produced from the reaction of a CH$_2\cdot$ radical with two CH$_3\cdot$ radicals. However, the probability of three species reacting simultaneously should be lower in the solid phase, and this reaction pathway may be negligible. The column density of propane molecules was not possible to determine, as its IR bands (e. g. 2960 cm$^{-1}$) overlap with vibration modes from a-C:H. The proposed reaction pathway for the formation of propane is provided below:

\begin{eqnarray}
\begin{split}
\label{Sch.C3H8}
      &\rm{C}_2\rm{H}_5 \cdot \thinspace + \thinspace \rm{CH}_3 \cdot \medspace \xrightarrow{} \thinspace \rm{C}_3\rm{H}_8\\
      &\rm{CH}_2 \cdot \thinspace + \thinspace 2 \thinspace \rm{CH}_3 \cdot \medspace \xrightarrow{} \thinspace \rm{C}_3\rm{H}_8
\end{split}
\end{eqnarray}\\

Ethyl radicals produced from methane dissociation may also react, forming butane molecules.

\begin{eqnarray}
\label{Sch.C4H10}
  \rm{C}_2\rm{H}_5 \cdot \thinspace + \thinspace \rm{C}_2\rm{H}_5 \cdot \medspace \xrightarrow{} \thinspace \rm{C}_4\rm{H}_{10}
\end{eqnarray}\\

C$_4$H$_{10}$ can also be produced from the rupture of propane molecules and the subsequent reaction with a methyl radical (Scheme \ref{Sch.C4H10.2}). However, as propane was found to be less abundant, this reaction is not expected to be influential in our experiments. Hence no butane was detected among the photoproducts in our experiments.

\begin{eqnarray}
\begin{split}
\label{Sch.C4H10.2}
  &\rm{C}_3\rm{H}_8 \thinspace + \thinspace h\nu \medspace \xrightarrow{} \thinspace \rm{C}_3\rm{H}_7 \cdot \medspace + \;\;\medspace \rm{H} \cdot\\
  &\rm{C}_3\rm{H}_7 \cdot \thinspace + \thinspace \rm{CH}_3 \cdot \medspace \xrightarrow{} \thinspace \rm{C}_4\rm{H}_{10}
\end{split}
\end{eqnarray}\\

We mentioned that chemical reactions between radicals formed by CH$_4$ ice irradiation lead to a complex branched network of C and H atoms, known as hydrogenated amorphous carbon, a-C:H \citep{Dartois2005}. The refractory a-C:H is the main constituent of carbonaceous dust grains observed in the ISM \citep{Dartois2005}. Subsequent irradiation leads to a reduction of C-H bonds in a-C:H, the C to H rate changes during the irradiation, and a less hydrogenated amorphous carbon is obtained.

\begin{eqnarray}
\label{Sch.aCH}
   \rm{x CH}_3 \cdot \thinspace + \thinspace \rm{y CH}_2 \cdot \thinspace + \thinspace \rm{z CH} \cdot \medspace \xrightarrow{} \thinspace \rm{a-C:H}
\end{eqnarray}\\

\subsection{Vacuum UV spectra}
\begin{figure} 
  \centering 
  \includegraphics[width=0.5\textwidth]{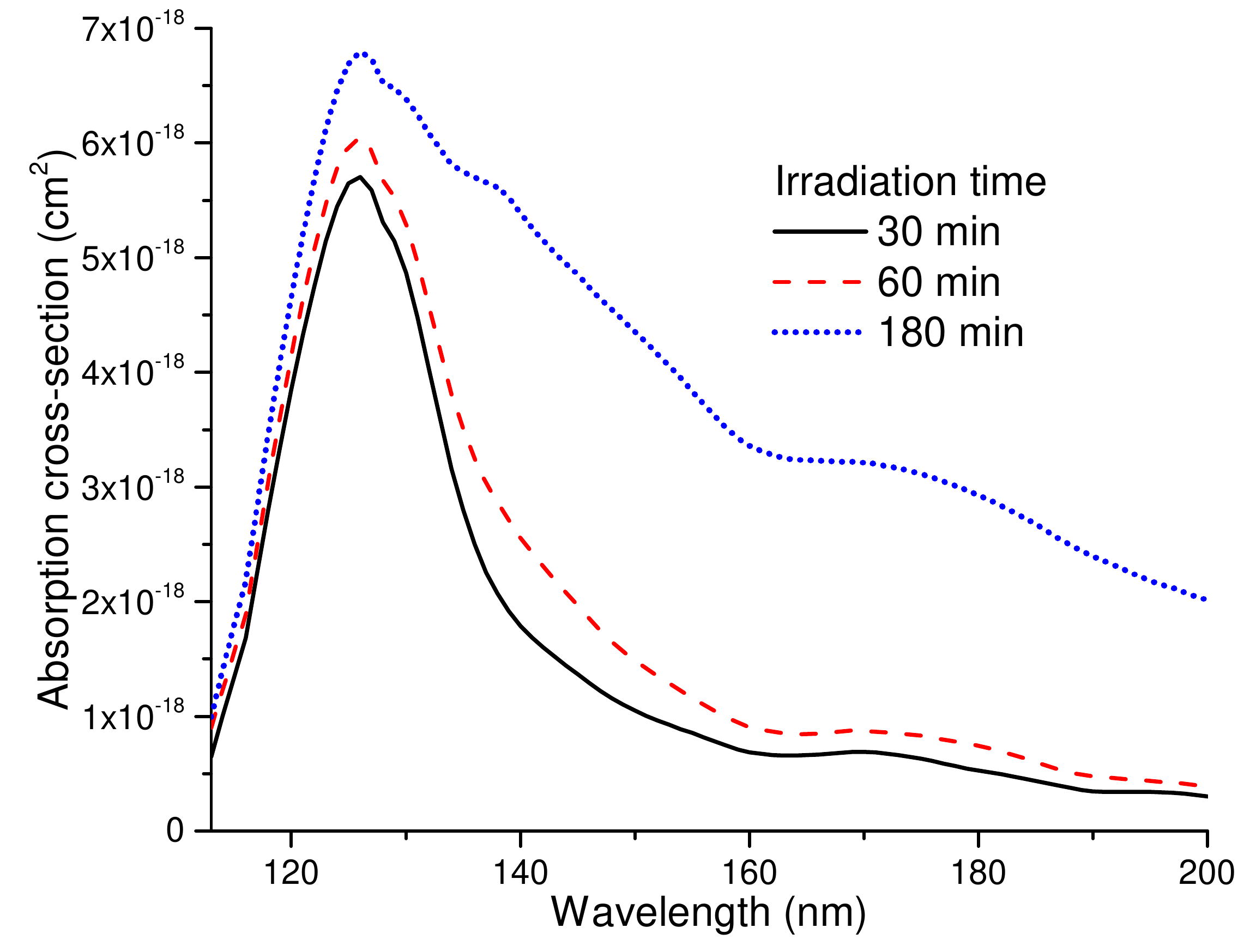}
  \caption{VUV absorption spectra of CH$_4$ ice. The bump appearing at 170 nm could be due to the presence of C$_2$H$_4$ molecules, see text.}
  \label{Fig.VUV}
\end{figure}

Fig. \ref{Fig.VUV} shows the VUV spectrum of methane ice. CH$_4$ has a strong absorption from the cut-off of the lamp (114 nm, 10.87 eV) to 133 nm (9.32 eV), and almost no absorption above 140 nm (8.86 eV) \citep{Wu2012ApJ...746..175W, Gus2014b}. During the irradiation steps, the formation of different photoproducts involving C-C bonds is responsible for the extended UV absorbance up to 180 nm (6.89 eV). These include C$_2$H$_6$, C$_3$H$_8$, the a-C:H residue, and probably C$_2$H$_2$ identified as a photodesorbing species in Sect. \ref{QMS}. C$_2$H$_4$, which was not identified by IR or QMS, could be responsible for the bump appearing between 160-180 nm (7.75-6.89 eV), according to its UV absorption spectrum \citep{BingMing2004, Young2018}.\\

\subsection{Photon-induced desorption}
\label{QMS}
\begin{figure*}
\centering
\includegraphics[width=\textwidth]{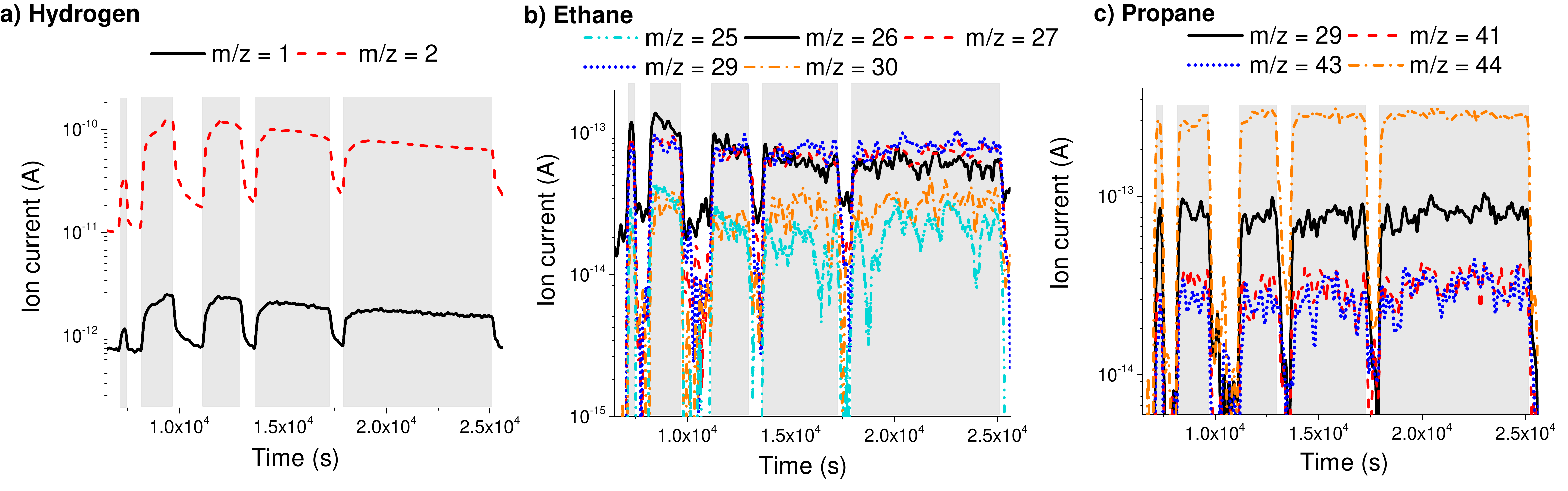}
\includegraphics[width=\textwidth]{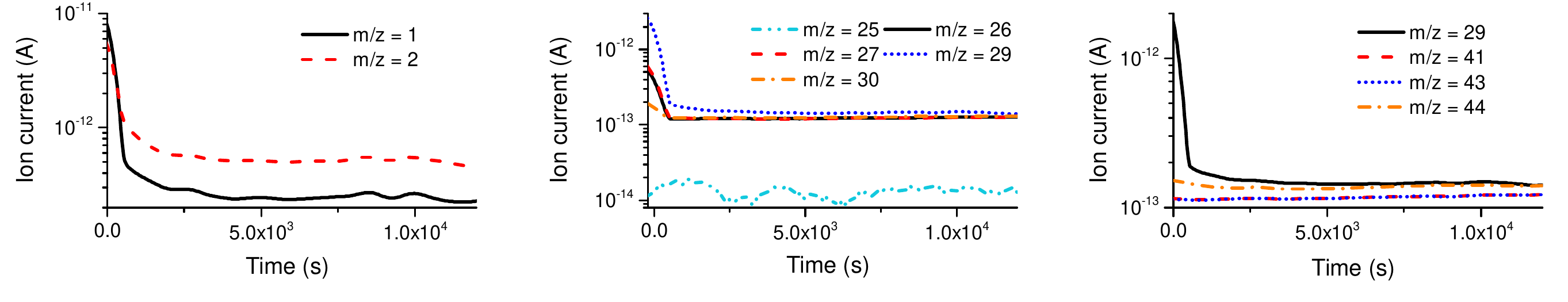}
\caption{Top: photodesorption of the three main photoproducts formed from UV-irradiation of CH$_4$, H$_2$, C$_2$H$_6$ and C$_3$H$_8$. Gray areas represent the time lapse when the lamp is turned on. Bottom: same $\frac{m}{z}$ fragments from a blank experiment where CH$_4$ ice was not irradiated. Note that the Y-axis is in a logarithmic scale.}
\label{Fig.PhotoCH4}
\end{figure*}

The QMS monitored the desorption of molecules from the ice sample. The time-dependent ion current from each molecular fragment served to detect the evolution of molecules in the gas-phase (see Sect. \ref{experimental_setup}). Fig. \ref{Fig.PhotoCH4}A, B, and C represent the ion current measured during the irradiation of pure CH$_4$ ice for the main $\frac{m}{z}$ fragments from H$_2$, C$_2$H$_6$, and C$_3$H$_8$, respectively.\\

H$\cdot$ radicals produced from methane dissociation can diffuse through the ice even at 8 K, thus forming hydrogen molecules that can escape from the ice to the gas-phase, allowing their detection by QMS for short irradiation times. The presence of H$_2$ background contamination intrinsic to UHV chambers and its possible release from turbomolecular pumps prevented the quantification of H$_2$ photodesorption. Additionally, the ion current generated by H$_2$ in the QMS is reduced for longer irradiation times as the column density of CH$_4$ ice diminishes.\\

\cite{Dupuy2017} reported CH$_4$ photodesorption from pure CH$_4$ ice for monochromatic photon energies above 9 eV, where the UV absorption of CH$_4$ becomes important \citep{Gus2014b}. They found photodesorption of methane to be around 2.3 $\times$ 10$^{-3}$ molecules per photon at Ly-$\alpha$ photon energy, and 2.2 $\times$ 10$^{-3}$ molecules per photon integrating over the \cite{Gredel1989ApJ...347..289G} photon distribution. However, CH$_4$ was not found to photodesorb in our experiments using the continuum emission MDHL, with a calculated upper limit of 1.7 $\times$ 10$^{-4}$ molecules per incident photon. This difference could be explained by the onset of CH$_4$ photodesorption starting at energies of about 9.2 eV (135 nm), with a large absorption cross-section of solid methane at Ly-$\alpha$, and even higher above 11 eV \citep{Wu2012ApJ...746..175W}, while there is almost no UV-absorption corresponding to the MDHL molecular emission range below 9.2 eV, i. e. a lower average absorption cross-section over the photon energy range covered by the MDHL, see \cite{Gus2014b}. Besides, when UV photons dissociate CH$_4$ ice molecules, H$\cdot$ atoms escape from the solid phase, leaving CH$_3\cdot$ radicals behind. Therefore, the backward reaction to produce CH$_4$ is inhibited, in particular at the surface of the ice, which may explain the absence of CH$_4$ photochemidesorption in our experiments with pure CH$_4$ ice. \cite{Gus2016} found that UV irradiation of pure CH$_3$OH ice gives rise to CH$_3\cdot$ and HCO species, among others. A likely explanation is that in \cite{Gus2016} experiments, CH$_3\cdot$ radicals react with HCO$\cdot$ and other species, allowing the formation of CH$_4$ molecules and their subsequent photochemidesorption when these reactions occur at the surface.\\

NH$\cdot$ and NH$_2\cdot$ radicals produced from UV irradiation of pure NH$_3$ ice were found to react mainly at temperatures well above 10 K \citep{MD2018}. Thus, the photoprocessing of the ice is limited to the production of radicals, and other photoproducts appear mainly during the warm-up phase. On the contrary, our experiments showed that radicals produced from UV irradiation of CH$_4$ are more reactive at low temperature, in agreement with other works \citep[e. g.][]{Lin2014}.\\

CH$_3\cdot$ radicals produced from CH$_4$ dissociation react in the ice, leading to an efficient ethane formation at the ice bulk and surface. QMS detected photon-induced desorption of the so-formed ethane molecules. The energy released during the formation of ethane from two methyl radicals overcomes the intermolecular interactions of the new molecule with the ice, which therefore photochemidesorbs as soon as it is formed (with a rate of $8\times10^{-4}$ molecules per incident photon\footnote{Values of the photon-induced desorption yields can vary by a factor of 2, estimated by the errors in the parameters in Eq. \ref{Eq.QMS}, see \cite{MD2016}}). A similar effect is observed for propane, which is, to our knowledge, the largest molecule found to photochemidesorb, with a rate of $2.4\times10^{-3}$ molecules per incident photon. The constant QMS ion current measured for ethane and propane molecules supports a photochemidesorption mechanism \citep{Gus2016, MD2015, MD2016}. The surface of the ice is renewed upon UV-irradiation due to formation of photoproducts and their eventual desorption if these species are formed at the ice surface. Radical formation and recombination also occurs in the ice bulk and become gradually exposed at the renewed surface due to desorption of the top monolayers during irradiation. The surface is thus enriched in ethane and propane molecules. The constant desorption rates of these species during irradiation indicates that only when they are formed directly on the surface, they can escape from the ice. Indeed, an ethane/propane molecule formed in the bulk that becomes later exposed to the surface will photodissociate rather than desorb when it absorbs a UV-photon.\\

Interestingly, Fig. \ref{Fig.PhotoCH4}B shows that $\frac{m}{z}$ = 25, and 26, decrease over the irradiation steps, while $\frac{m}{z}$ = 27, 28, 29, and 30 remain constant, suggesting that a second contribution is present within $\frac{m}{z}$ = 25, and 26. C$_2$H$_2$ photodesorption may be responsible for this behaviour, since $\frac{m}{z}$ = 26 is the most intense molecular fragment of acetylene, followed by $\frac{m}{z}$ = 25, and almost no $\frac{m}{z}$ = 24 is produced. Acetylene can be produced from dehydrogenation of ethane molecules, although ethane photochemidesorption would inhibit acetylene formation in the surface, and therefore its photodesorption. C$_2$H$_2$ can also be formed from the reaction between two CH$\cdot$ radicals produced directly from CH$_4$ dissociation, as shown in Scheme \ref{Sch.C2H2}. \cite{Okabe1978} reports a yield value of 0.06 with respect to the production of CH$_3\cdot$ radicals for the gas-phase using Ly-$\alpha$ photons. Recently, \cite{Lin2014} measured the threshold energy of this reaction in CH$_4$ ice at 3 K, obtaining a value of 9.2 eV (134.8 nm). Therefore, the continuum emission of our MDHL may favour this reaction.\\

\begin{eqnarray}
\begin{split}
\label{Sch.C2H2}
  &2 \medspace CH \cdot \medspace \xrightarrow{} \thinspace C_2H_2\\
\end{split}
\end{eqnarray}\\

\subsection{Thermal desorption}

\begin{figure} 
  \centering 
  \includegraphics[width=0.5\textwidth]{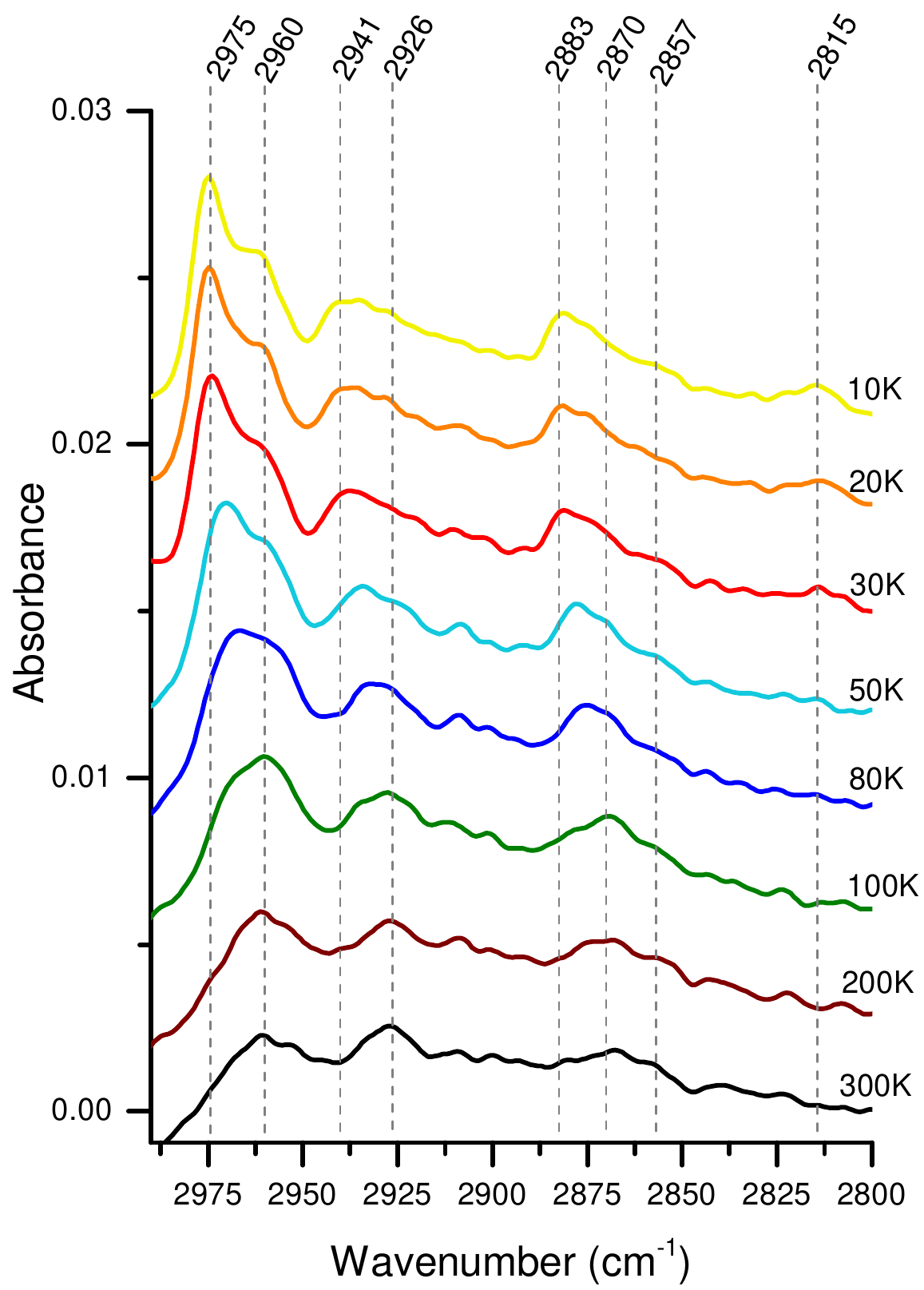}
  \caption{IR spectra collected during warm-up of irradiated CH$_4$ ice at different temperatures. Spectra were offset for clarity.}
  \label{Fig.TPD-IR}
\end{figure}

After irradiation, CH$_4$ ice samples were warmed up at a rate of 1 K/min. IR spectra were collected during the temperature programmed desorption (TPD). Methane bands at 3009, 2815 and 1300 cm$^{-1}$ disappear at a temperature lower than 40 K, while photoproducts (mainly ethane, propane and hydrogenated amorphous carbon), will remain in the ice at higher temperatures. Fig. \ref{Fig.TPD-IR} shows the IR spectra measured during the TPD at different temperatures. Bands identified with ethane (2975, 2941 and 2883 cm$^{-1}$) disappear gradually as the temperature increases, and they are not observed above 70 K. Meanwhile, IR bands at 2960 cm$^{-1}$, 2926 cm$^{-1}$, 2870 cm$^{-1}$ and 2857 cm$^{-1}$, related to different stretching modes of a-C:H \citep{Dartois2005}, are still clearly observable at 300 K.\\

\begin{table} 
    \caption[]{Binding energy, in K, of desorbing species during the warm-up of the samples.}
    \label{Table.binding.energies}
    \begin{center}
    \begin{tabular}{cccccccc}
Species & T$_{des}$(K) & E$_b$(K)&&Species & T$_{des}$(K) & E$_b$(K)\\
\noalign{\smallskip}
\cline{1-3}
\cline{5-7}
\noalign{\smallskip}
\noalign{\smallskip}
CH$_4$ & 38.6  &1193    &&$^{13}$CH$_4$     &38.7   &1196\\
\noalign{\smallskip}
C$_2$H$_6$  &66.1 &2042 &&$^{13}$C$_2$H$_6$ &67.0   &2070\\
\noalign{\smallskip}
C$_3$H$_8$  &83.5 &2580 &&$^{13}$C$_3$H$_8$ &85.0   &2626\\
\noalign{\smallskip}
\hline
\noalign{\smallskip}
\end{tabular}\\
\end{center}
\end{table}

\begin{figure*} 
  \centering 
\includegraphics[width=\textwidth]{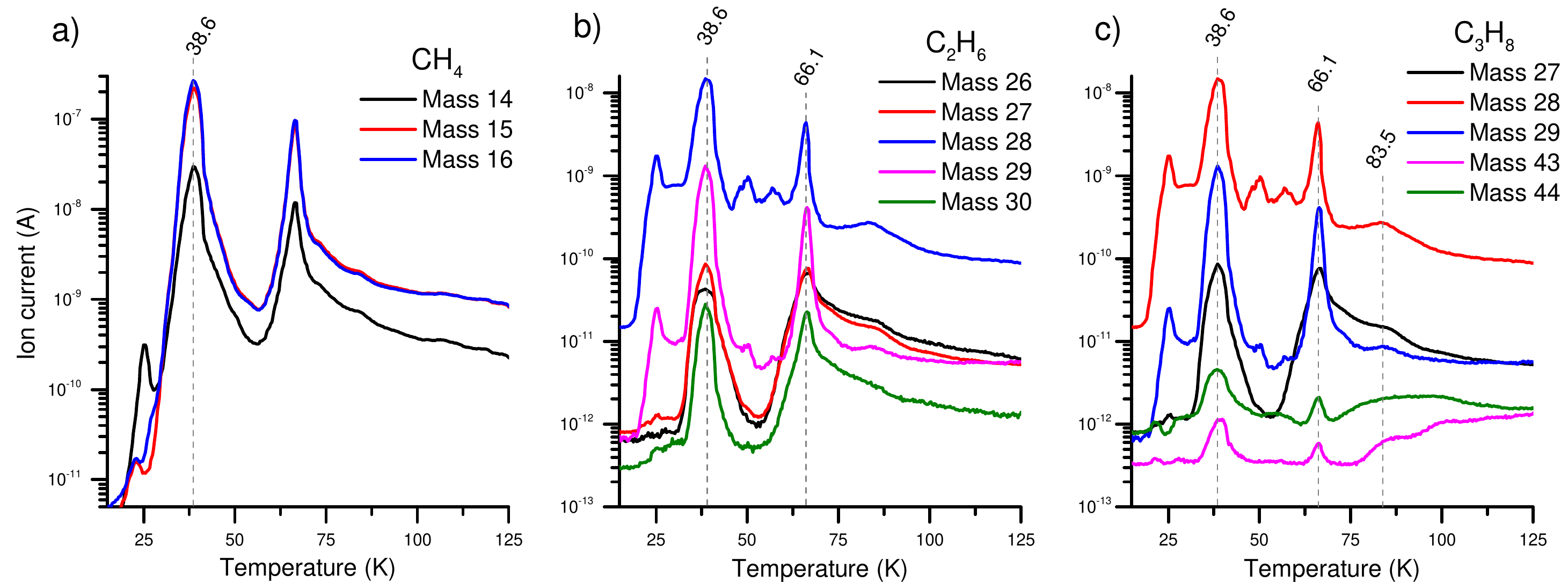}
\includegraphics[width=\textwidth]{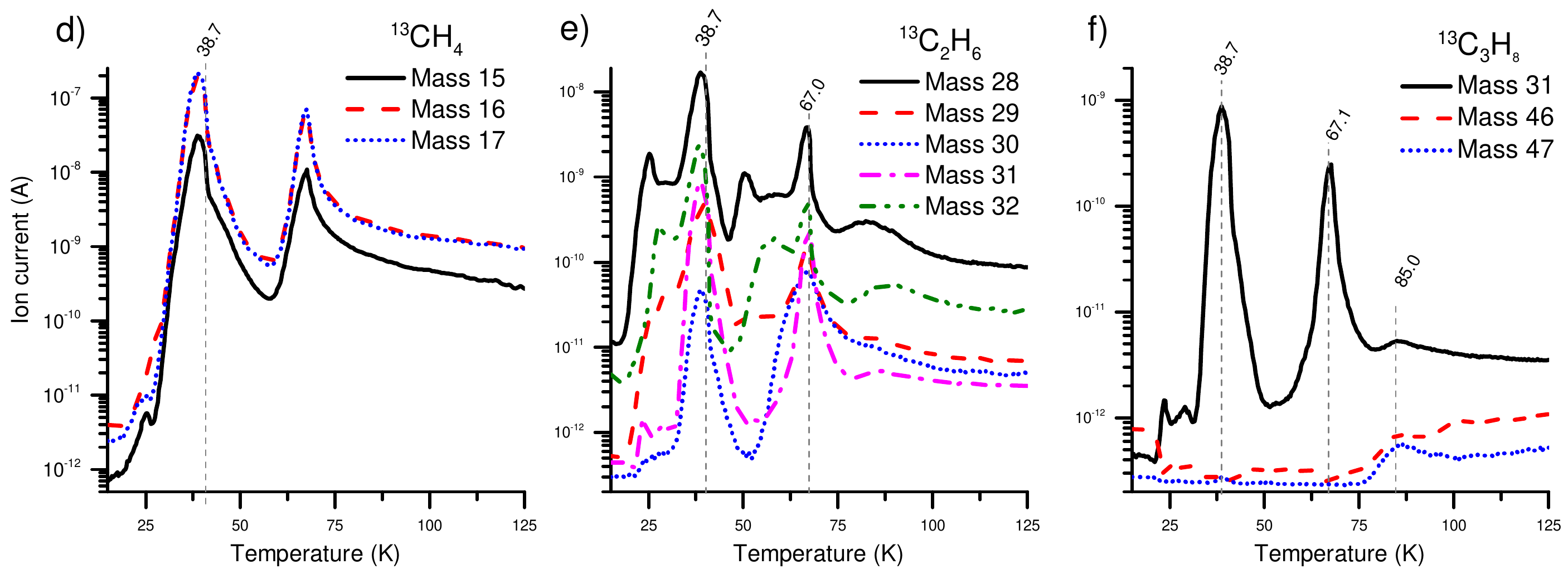}
  \caption{TPD curves for the main species present in the ice after the irradiation period for pure CH$_4$ ice (top) and for pure $^{13}$CH$_4$ ice (bottom). Note that the Y-axis is in a logarithmic scale.}
  \label{Fig.TPDCH4}
\end{figure*}

The ion current of different $\frac{m}{z}$ fragments during the TPD process is represented in Fig. \ref{Fig.TPDCH4}. CH$_4$ molecules (Fig. \ref{Fig.TPDCH4}A) desorb thermally from the irradiated CH$_4$ ice at 38.6 K. A second peak that appears near 66.5 K is related to the thermal desorption of methane ice that accreted outside the sample substrate in cold areas of the cold finger. Indeed, this peak disappears if the cold finger of the cryostat is warmed up at the same rate with the thermal resistance and the cryostat off (\textit{natural} TPD). The thermal desorption of C$_2$H$_6$ occurs at 66.1 K. Two clear peaks are also observed in Fig. \ref{Fig.TPDCH4}B. The one at 38.6 K represents the ethane molecules co-desorbing with CH$_4$, with $\frac{m}{z}$ = 30 for the molecular ion, C$_2$H$_6^{+}$, and other fragments at $\frac{m}{z}$ = 29, 28, 27, 26 and 25 with relative intensities similar to those in the NIST database for ethane. As CH$_4$ is the primary molecule present in the ice, its thermal desorption can drag other molecules. But other ethane molecules will remain in the ice until 66.1 K, where C$_2$H$_6$ molecules desorb from pure ethane ice (in agreement with \cite{Hudson2009}). Propane co-desorption with methane and ethane molecules is observed at 38.6 K and 66.1 K, respectively. The third peak, appearing at 83.5 K, belongs to the propane ice thermal desorption. Following \cite{Luna2017}, the binding energies of these species are obtained multiplying the peak temperatures by a factor of 30.9. These values for CH$_4$, C$_2$H$_6$ and C$_3$H$_8$ and their homologous $^{13}$C isotopologues are shown in Table \ref{Table.binding.energies}.\\

To further confirm the results obtained for CH$_4$ ice, experiments were repeated using $^{13}$CH$_4$. The ion current of different $\frac{m}{z}$ fragments is represented in Fig. \ref{Fig.TPDCH4} (bottom part) for a TPD from a $^{13}$CH$_4$ ice sample. $^{13}$CH$_4$ desorbs thermally at 38.7 K. $^{13}$C$_2$H$_6$ co-desorbs with methane, while its pure thermal desorption occurs at 67.0 K. Both processes are similar to the ones found for the CH$_4$ ice. Propane co-desorption with CH$_4$ and C$_2$H$_6$ at 38.6 K and 66.1 K in the CH$_4$ experiment, monitored by $\frac{m}{z}$ = 43 and 44, were not confirmed by their corresponding fragments $\frac{m}{z}$ = 46 and 47 in the $^{13}$CH$_4$ experiment. While $\frac{m}{z}$ = 44 could suffer from CO$_2$ contamination, the $\frac{m}{z}$ = 43 desorption could not be identified with a contaminant. Nevertheless, the desorption of C$_3$H$_8$ at 83.5 K was reproduced in the $^{13}$CH$_4$ experiment as a peak appearing at 85.0 K (Fig. \ref{Fig.TPDCH4}F). Thermal desorption of $^{13}$CH$_4$, $^{13}$C$_2$H$_6$ and $^{13}$C$_3$H$_8$ takes place at slightly higher temperatures when compared to the homologous species with $^{12}$C, as a consequence of their larger molecular mass.\\

In this work, $^{13}$CH$_4$ was only employed to rule out any potential source of carbon contamination. Some processes were not quantified, such as propane formation or photochemidesorption of species in the $^{13}$CH$_4$ experiment. Despite this, similar trends regarding photochemistry and photodesorption were obtained.\\


\section{Astrophysical implications}
In the interior of dense clouds, the dust temperature is too low to induce thermal processes, and the interstellar radiation UV field cannot penetrate deep into the cloud. Therefore, secondary UV-photons drive the photolytic processes in these regions (see Sect. \ref{Introduction}). This work reports measurements of the UV photon-induced processes over a pure CH$_4$ ice, as a first approximation to a more realistic scenario, in line with previous works that addressed the photon-induced behaviour of pure ice components. Since CH$_4$ is thought to be mainly formed from successive hydrogenation of C atoms \citep{Oberg2008}, methane might be present in a H$_2$O-rich environment. Intermolecular forces binding CH$_4$ molecules in the ice will obviously depend on the ice mantle composition and this will also affect strongly its photodesorption and photochemistry. For instance, contrary to the case of pure CH$_4$ ice, reported in this paper, CH$_4$ formed by methanol or ethanol ice irradiation is found to photochemidesorb \citep{Gus2016,MD2016}.\\

\cite{Boogert1996, d'hendecourt1996A&A...315L.365D} reported methane abundance in dense interstellar clouds up to 1.9\% relative to water. In addition, \cite{d'hendecourt1996A&A...315L.365D} estimated a column density of 4.3$\times$10$^{17}$ cm$^{-2}$ toward RAFGL 7009S, a circumstellar environment, representing a 4\% abundance relative to water ice. \cite{Oberg2008} analyzed CH$_4$ abundances toward different low-mass young stellar objects, finding an average value of 5.8\% abundance relative to water. CH$_4$ is expected to play a role in the production of COMs in the ISM. From pure CH$_4$ ice irradiation, larger molecules are formed, and eventually ejected to the gas-phase, such as ethane and propane. Complementary experiments are required to gain a better understanding of the behaviour of CH$_4$ in different ice mixtures, such as H$_2$O:CH$_4$ \citep{MD2016} or CO:CH$_4$ mixtures. In the H$_2$O:CH$_4$ ice irradiation, \cite{MD2016} observed the photon-induced desorption of formaldehyde. \cite{Oberg2010} found that the destruction cross-section of CH$_4$ molecules is increased in a water-rich environment.\\

Photochemidesorption of C$_2$H$_6$ was found to be about 8$\times$10$^{-4}$ molecules per incident photon, while that of C$_3$H$_8$ is 2.4$\times$10$^{-3}$ molecules per incident photon. As pure CH$_4$ ice may not be present within interstellar conditions, these values can be taken as upper limits to explain the gas-phase abundance of ethane and propane molecules. On the other hand, pure CH$_4$ is abundant in transneptunian objects (TNOs) such as Triton or Pluto \citep[and references therein]{Owen1993} and may be a common ice component in outer regions of protoplanetary disks. The composition of Pluto's atmosphere contains not only N$_2$ and CH$_4$, but also C$_2$H$_2$, C$_2$H$_4$, and C$_2$H$_6$ \citep{Young2018}. The surface temperature is high enough to allow thermal desorption of CH$_4$, explaining its presence in the gas-phase, while other species require the presence of alternative pathways. Our experiments show that photon-induced processes, as discussed in \cite{Wong2017}, can contribute to the processes enriching the gas-phase abundance of C$_2$H$_6$ and C$_2$H$_2$ species, while C$_2$H$_4$ was not detected in the gas-phase. These results qualitatively agree with Pluto observations reported by \cite{Young2018}, which measured C$_2$H$_6$ to be the most abundant hydrocarbon (apart from CH$_4$), followed by C$_2$H$_2$ and C$_2$H$_4$.\\

Our results can also be extended to the discussion of the evolution of carbon-enriched comets, which contain relatively large abundances of methane molecules, up to 1\% relative to water \citep{Mumma2011}. In these objects, photolysis of CH$_4$ gives rise to ethane and acetylene molecules, with cometary abundances ranging from 0.1-2\% and 0.09-0.5\%, respectively, which were also detected in our experiments. A full study of the evolution with more realistic CH$_4$/H$_2$O cometary ice analog mixtures is required to better constrain the photochemistry in these objects.


\section{Conclusions}
\cite{Lin2014} and \cite{Lo2015MNRAS.451..159L} irradiated CH$_4$ samples, both pure and dispersed in a neon matrix, at 3 K, with monochromatic UV photons at wavelengths ranging from Ly-$\alpha$ to 190 nm. In addition to other photoproducts, they reported the formation of C$_2$H$_6$, C$_2$H$_4$ and C$_2$H$_2$ molecules as well as CH$_3\cdot$ and CH$\cdot$ radicals. They found CH$_x\cdot$ radicals to react even at 3 K in a CH$_4$:Ne = 1:10.000 matrix. In our experiments, CH$_x\cdot$ radicals were not detected, as their lifetime is not high enough at 8 K. Their irradiation experiments at Ly-$\alpha$ wavelengths showed similar abundances to our experiments, C$_2$H$_6$ being the most abundant photoproduct, followed by C$_2$H$_2$ and C$_2$H$_4$. However, since the abundances depend on the wavelength used for irradiation, a different emission spectrum may change the photon-induced desorption processes found in our experiments.\\

In addition to the monochromatic irradiation of methane samples, \cite{Bossa2015PCCP...1717346B} used a continuum emission hydrogen lamp, similar to our experiments. They deposited samples at 20 K, thus having a phase II methane instead of an amorphous one \citep{Hudson2015}. They reported the formation of ethane, ethylene and acetylene, but, as in our experiments, radical species were not detected.\\

Similar photoproducts were found by \cite{Abplanalp2018PCCP...20.5435A}, who irradiated methane ices at 5.5 K with keV electrons, monochromatic Ly-$\alpha$ photons and continuum emission photons (from 112.7-169.8 nm, or 11.0-7.3 eV) using a deuterium lamp. They reported the presence of radical species and small hydrocarbons measured in the ice by infrared spectroscopy. Additionally, they reported the detection of larger hydrocarbon chains and cyclic species during the warm-up of the samples by time-of-flight mass spectrometry analysis.\\

\cite{Baratta2002} reported experiments of a pure CH$_4$ ice irradiated by 30 keV He$^{+}$ ions and UV photons. They measured the column density of methane, ethane and propane after different irradiation doses. \cite{deBarros2011} compared the effects of pure CH$_4$ ice processing using different irradiation sources: oxygen MeV ions, protons, $\alpha$-particles and electrons. In this work, we have provided experiments regarding UV-irradiation (from 7.6 eV to 10.8 eV) of pure CH$_4$ ice aiming to study the photon-induced desorption of photoproducts. Ethane molecules were found to account for 54\% of the dissociated methane molecules at the beginning of the irradiation. Other photoproducts, such as C$_3$H$_8$ and the progressive formation of a-C:H, account for the rest of C atoms from methane dissociation, although the proportions varied during the irradiation. Propane was not quantified, as its main bands overlap with those of a-C:H, which is formed from the beginning of the irradiation, in agreement with \cite{Dartois2005}.\\

\cite{Gerakines1996} measured the photolysis and formation cross-section of CH$_4$ and C$_2$H$_6$, respectively. They found $\sigma_{des}$ of methane to be 7.2$\times$10$^{-19}$ cm$^{2}$, while our experiments showed a value of 2.2$\times$10$^{-18}$ cm$^{2}$. The following reasons could explain our higher $\sigma_{des}$. First, \cite{Gerakines1996} calculated their value as a function of the incident photon dose, while we took into account the absorption from the ice, thus removing the contribution of low energy UV-photons which are not absorbed by the CH$_4$ ice sample. Additionally, the total flux was estimated in our experiments by in-situ measurements with a Ni-mesh (see \cite{Cristobal2019}) whereas \cite{Gerakines1996} assumed a constant mean flux of 10$^{15}$ photons cm$^{-2}$ s$^{-1}$. Finally, the poorer vacuum conditions in \cite{Gerakines1996} setup could contribute to reduce the photodissociation of CH$_4$ due to background water accretion. $\sigma_{for}$ of ethane found in our experiments (1.2$\times$10$^{-18}$ cm$^{2}$) is also larger than the one they reported (3.2$\times$10$^{-19}$ cm$^{2}$).\\

Photodesorption mechanisms were studied for CH$_4$ and its daughter molecules. \cite{Dupuy2017} found CH$_4$ photodesorption for monochromatic energies above 9.1 eV. However, in agreement with \cite{Gus2014b} and \cite{MD2016}, no evidence of a significant CH$_4$ photon-induced desorption was found in our work using the MDHL. Indeed, in our experiments, methane molecules are prone to dissociate rather than photodesorb. Recombination of radicals produce ethane and propane molecules, which were found to photochemidesorb. The excess energy released from the formation of photoproducts contributes to their desorption on the top monolayer. The constant ion current obtained for ethane and propane molecules during CH$_4$ ice irradiation supported the photochemidesorption mechanism, as explained in Sect. \ref{QMS}. Additionally, C$_2$H$_2$ was found to photodesorb, producing a difference in the ion current of $\frac{m}{z}$ = 25, and 26, when compared to other fragments of the hydrocarbons detected.\\

\section*{acknowledgements}
The Spanish Ministry of Science, Innovation and Universities supported this research under grant number AYA2017-85322-R (AEI/FEDER, UE), PhD fellowship FPU-17/03172 and MDM-2017-0737 Unidad de Excelencia "Mar\'{i}a de Maeztu"-- Centro de Astrobiolog\'{i}a (CSIC-INTA). G. A. Cruz-Diaz was supported by the National Aeronautics and Space Administration through the NASA Astrobiology Institute under Cooperative Agreement Notice NNH13ZDA017C issued through the Science Mission Directorate.



\bibliographystyle{mnras}
\bibliography{bibliography} 

\begin{thebibliography}{}
\makeatletter
\relax
\def\mn@urlcharsother{\let\do\@makeother \do\$\do\&\do\#\do\^\do\_\do\%\do\~}
\def\mn@doi{\begingroup\mn@urlcharsother \@ifnextchar [ {\mn@doi@}
  {\mn@doi@[]}}
\def\mn@doi@[#1]#2{\def\@tempa{#1}\ifx\@tempa\@empty \href
  {http://dx.doi.org/#2} {doi:#2}\else \href {http://dx.doi.org/#2} {#1}\fi
  \endgroup}
\def\mn@eprint#1#2{\mn@eprint@#1:#2::\@nil}
\def\mn@eprint@arXiv#1{\href {http://arxiv.org/abs/#1} {{\tt arXiv:#1}}}
\def\mn@eprint@dblp#1{\href {http://dblp.uni-trier.de/rec/bibtex/#1.xml}
  {dblp:#1}}
\def\mn@eprint@#1:#2:#3:#4\@nil{\def\@tempa {#1}\def\@tempb {#2}\def\@tempc
  {#3}\ifx \@tempc \@empty \let \@tempc \@tempb \let \@tempb \@tempa \fi \ifx
  \@tempb \@empty \def\@tempb {arXiv}\fi \@ifundefined
  {mn@eprint@\@tempb}{\@tempb:\@tempc}{\expandafter \expandafter \csname
  mn@eprint@\@tempb\endcsname \expandafter{\@tempc}}}

\bibitem[\protect\citeauthoryear{{Abplanalp}, {Jones}  \& {Kaiser}}{{Abplanalp}
  et~al.}{2018}]{Abplanalp2018PCCP...20.5435A}
{Abplanalp} M.~J.,  {Jones} B.~M.,   {Kaiser} R.~I.,  2018, \mn@doi [Physical
  Chemistry Chemical Physics (Incorporating Faraday Transactions)]
  {10.1039/C7CP05882A}, \href
  {https://ui.adsabs.harvard.edu/abs/2018PCCP...20.5435A} {20, 5435}

\bibitem[\protect\citeauthoryear{{Andersson} \& {van Dishoeck}}{{Andersson} \&
  {van Dishoeck}}{2008}]{Andersson2008}
{Andersson} S.,  {van Dishoeck} E.~F.,  2008, \mn@doi [\aap]
  {10.1051/0004-6361:200810374}, \href
  {https://ui.adsabs.harvard.edu/abs/2008A&A...491..907A} {491, 907}

\bibitem[\protect\citeauthoryear{{Baratta}, {Leto}  \& {Palumbo}}{{Baratta}
  et~al.}{2002}]{Baratta2002}
{Baratta} G.~A.,  {Leto} G.,   {Palumbo} M.~E.,  2002, \mn@doi [\aap]
  {10.1051/0004-6361:20011835}, \href
  {https://ui.adsabs.harvard.edu/abs/2002A%26A...384..343B} {384, 343}

\bibitem[\protect\citeauthoryear{{Bennett}, {Jamieson}, {Osamura}  \&
  {Kaiser}}{{Bennett} et~al.}{2006}]{Bennett2006}
{Bennett} C.~J.,  {Jamieson} C.~S.,  {Osamura} Y.,   {Kaiser} R.~I.,  2006,
  \mn@doi [\apj] {10.1086/508561}, \href
  {https://ui.adsabs.harvard.edu/\#abs/2006ApJ...653..792B} {653, 792}

\bibitem[\protect\citeauthoryear{{Bertin} et~al.,}{{Bertin}
  et~al.}{2016}]{Bertin2016}
{Bertin} M.,  et~al., 2016, \mn@doi [\apjl] {10.3847/2041-8205/817/2/L12},
  \href {https://ui.adsabs.harvard.edu/abs/2016ApJ...817L..12B} {817, L12}

\bibitem[\protect\citeauthoryear{{Boogert} et~al.,}{{Boogert}
  et~al.}{1996}]{Boogert1996}
{Boogert} A.~C.~A.,  et~al., 1996, \aap, \href
  {https://ui.adsabs.harvard.edu/\#abs/1996A&A...315L.377B} {315, L377}

\bibitem[\protect\citeauthoryear{{Boogert}, {Gerakines}  \&
  {Whittet}}{{Boogert} et~al.}{2015}]{Boogert2015ARA&A..53..541B}
{Boogert} A.~C.~A.,  {Gerakines} P.~A.,   {Whittet} D. C.~B.,  2015, \mn@doi
  [\araa] {10.1146/annurev-astro-082214-122348}, \href
  {https://ui.adsabs.harvard.edu/abs/2015ARA&A..53..541B} {53, 541}

\bibitem[\protect\citeauthoryear{{Bossa}, {Paardekooper}, {Isokoski}  \&
  {Linnartz}}{{Bossa} et~al.}{2015}]{Bossa2015PCCP...1717346B}
{Bossa} J.~B.,  {Paardekooper} D.~M.,  {Isokoski} K.,   {Linnartz} H.,  2015,
  \mn@doi [Physical Chemistry Chemical Physics (Incorporating Faraday
  Transactions)] {10.1039/C5CP00578G}, \href
  {https://ui.adsabs.harvard.edu/abs/2015PCCP...1717346B} {17, 17346}

\bibitem[\protect\citeauthoryear{{Cecchi-Pestellini} \&
  {Aiello}}{{Cecchi-Pestellini} \& {Aiello}}{1992}]{Cecchi1992}
{Cecchi-Pestellini} C.,  {Aiello} S.,  1992, \mn@doi [\mnras]
  {10.1093/mnras/258.1.125}, \href
  {https://ui.adsabs.harvard.edu/#abs/1992MNRAS.258..125C} {258, 125}

\bibitem[\protect\citeauthoryear{{Cruikshank}, {Roush}, {Owen}, {Geballe}, {de
  Bergh}, {Schmitt}, {Brown}  \& {Bartholomew}}{{Cruikshank}
  et~al.}{1993}]{Cruikshank1993}
{Cruikshank} D.~P.,  {Roush} T.~L.,  {Owen} T.~C.,  {Geballe} T.~R.,  {de
  Bergh} C.,  {Schmitt} B.,  {Brown} R.~H.,   {Bartholomew} M.~J.,  1993,
  \mn@doi [Science] {10.1126/science.261.5122.742}, \href
  {https://ui.adsabs.harvard.edu/\#abs/1993Sci...261..742C} {261, 742}

\bibitem[\protect\citeauthoryear{{Cruz-Diaz}, {Mu{\~n}oz Caro}, {Chen}  \&
  {Yih}}{{Cruz-Diaz} et~al.}{2014}]{Gus2014b}
{Cruz-Diaz} G.~A.,  {Mu{\~n}oz Caro} G.~M.,  {Chen} Y.~J.,   {Yih} T.~S.,
  2014, \mn@doi [\aap] {10.1051/0004-6361/201322621}, \href
  {https://ui.adsabs.harvard.edu/#abs/2014A&A...562A.120C} {562, A120}

\bibitem[\protect\citeauthoryear{{Cruz-Diaz}, {Mart{\'\i}n-Dom{\'e}nech},
  {Mu{\~n}oz Caro}  \& {Chen}}{{Cruz-Diaz} et~al.}{2016}]{Gus2016}
{Cruz-Diaz} G.~A.,  {Mart{\'\i}n-Dom{\'e}nech} R.,  {Mu{\~n}oz Caro} G.~M.,
  {Chen} Y.~J.,  2016, \mn@doi [\aap] {10.1051/0004-6361/201526761}, \href
  {https://ui.adsabs.harvard.edu/#abs/2016A&A...592A..68C} {592, A68}

\bibitem[\protect\citeauthoryear{{Dartois}}{{Dartois}}{2005}]{Dartois2005SSRv..119..293D}
{Dartois} E.,  2005, \mn@doi [\ssr] {10.1007/s11214-005-8059-9}, \href
  {https://ui.adsabs.harvard.edu/abs/2005SSRv..119..293D} {119, 293}

\bibitem[\protect\citeauthoryear{{Dartois}, {Mu{\~n}oz Caro}, {Deboffle},
  {Montagnac}  \& {d'Hendecourt}}{{Dartois} et~al.}{2005}]{Dartois2005}
{Dartois} E.,  {Mu{\~n}oz Caro} G.~M.,  {Deboffle} D.,  {Montagnac} G.,
  {d'Hendecourt} L.,  2005, \mn@doi [\aap] {10.1051/0004-6361:20042094}, \href
  {https://ui.adsabs.harvard.edu/\#abs/2005A&A...432..895D} {432, 895}

\bibitem[\protect\citeauthoryear{{Dupuy} et~al.,}{{Dupuy}
  et~al.}{2017}]{Dupuy2017}
{Dupuy} R.,  et~al., 2017, \mn@doi [\aap] {10.1051/0004-6361/201730772}, \href
  {http://adsabs.harvard.edu/abs/2017A%26A...603A..61D} {603, A61}

\bibitem[\protect\citeauthoryear{{Fayolle} et~al.,}{{Fayolle}
  et~al.}{2013}]{Fayolle2013}
{Fayolle} E.~C.,  et~al., 2013, \mn@doi [\aap] {10.1051/0004-6361/201321533},
  \href {https://ui.adsabs.harvard.edu/#abs/2013A&A...556A.122F} {556, A122}

\bibitem[\protect\citeauthoryear{{Fillion} et~al.,}{{Fillion}
  et~al.}{2014}]{Fillion2014}
{Fillion} J.-H.,  et~al., 2014, \mn@doi [Faraday Discussions]
  {10.1039/C3FD00129F}, \href
  {https://ui.adsabs.harvard.edu/#abs/2014FaDi..168..533F} {168, 533}

\bibitem[\protect\citeauthoryear{{Gerakines}, {Schutte}  \&
  {Ehrenfreund}}{{Gerakines} et~al.}{1996}]{Gerakines1996}
{Gerakines} P.~A.,  {Schutte} W.~A.,   {Ehrenfreund} P.,  1996, \aap, \href
  {https://ui.adsabs.harvard.edu/#abs/1996A&A...312..289G} {312, 289}

\bibitem[\protect\citeauthoryear{{Gonz{\'a}lez D{\'{\i}}az}, {Carrascosa de
  Lucas}, {Aparicio}, {Mu{\~n}oz Caro}, {Sie}, {Hsiao}, {Cazaux}  \&
  {Chen}}{{Gonz{\'a}lez D{\'{\i}}az} et~al.}{2019}]{Cristobal2019}
{Gonz{\'a}lez D{\'{\i}}az} C.,  {Carrascosa de Lucas} H.,  {Aparicio} S.,
  {Mu{\~n}oz Caro} G.~M.,  {Sie} N.-E.,  {Hsiao} L.-C.,  {Cazaux} S.,   {Chen}
  Y.-J.,  2019, \mn@doi [\mnras] {10.1093/mnras/stz1223}, \href
  {https://ui.adsabs.harvard.edu/abs/2019MNRAS.486.5519G} {486, 5519}

\bibitem[\protect\citeauthoryear{{Gredel}, {Lepp}, {Dalgarno}  \&
  {Herbst}}{{Gredel} et~al.}{1989}]{Gredel1989ApJ...347..289G}
{Gredel} R.,  {Lepp} S.,  {Dalgarno} A.,   {Herbst} E.,  1989, \mn@doi [\apj]
  {10.1086/168117}, \href
  {https://ui.adsabs.harvard.edu/abs/1989ApJ...347..289G} {347, 289}

\bibitem[\protect\citeauthoryear{{Grundy} et~al.,}{{Grundy}
  et~al.}{2016}]{Grundy2016}
{Grundy} W.~M.,  et~al., 2016, \mn@doi [Science] {10.1126/science.aad9189},
  \href {https://ui.adsabs.harvard.edu/\#abs/2016Sci...351.9189G} {351,
  aad9189}

\bibitem[\protect\citeauthoryear{{Hudson}, {Moore}  \& {Raines}}{{Hudson}
  et~al.}{2009}]{Hudson2009}
{Hudson} R.~L.,  {Moore} M.~H.,   {Raines} L.~L.,  2009, \mn@doi [\icarus]
  {10.1016/j.icarus.2009.06.026}, \href
  {https://ui.adsabs.harvard.edu/\#abs/2009Icar..203..677H} {203, 677}

\bibitem[\protect\citeauthoryear{{Hudson}, {Gerakines}  \& {Loeffler}}{{Hudson}
  et~al.}{2015}]{Hudson2015}
{Hudson} R.~L.,  {Gerakines} P.~A.,   {Loeffler} M.~J.,  2015, \mn@doi
  [Physical Chemistry Chemical Physics (Incorporating Faraday Transactions)]
  {10.1039/C5CP00975H}, \href
  {https://ui.adsabs.harvard.edu/abs/2015PCCP...1712545H} {17, 12545}

\bibitem[\protect\citeauthoryear{{Kaiser} \& {Roessler}}{{Kaiser} \&
  {Roessler}}{1998}]{Kaiser1998}
{Kaiser} R.~I.,  {Roessler} K.,  1998, \mn@doi [\apj] {10.1086/306001}, \href
  {https://ui.adsabs.harvard.edu/\#abs/1998ApJ...503..959K} {503, 959}

\bibitem[\protect\citeauthoryear{{Lacy}, {Carr}, {Evans}, {Baas}, {Achtermann}
  \& {Arens}}{{Lacy} et~al.}{1991}]{Lacy1991}
{Lacy} J.~H.,  {Carr} J.~S.,  {Evans} Neal~J. I.,  {Baas} F.,  {Achtermann}
  J.~M.,   {Arens} J.~F.,  1991, \mn@doi [\apj] {10.1086/170304}, \href
  {https://ui.adsabs.harvard.edu/\#abs/1991ApJ...376..556L} {376, 556}

\bibitem[\protect\citeauthoryear{Lin, Lo, Lu, Chou, Peng, Cheng  \&
  Ogilvie}{Lin et~al.}{2014}]{Lin2014}
Lin M.-Y.,  Lo J.-I.,  Lu H.-C.,  Chou S.-L.,  Peng Y.-C.,  Cheng B.-M.,
  Ogilvie J.~F.,  2014, \mn@doi [The Journal of Physical Chemistry A]
  {10.1021/jp502637r}, 118, 3438

\bibitem[\protect\citeauthoryear{{Lo}, {Lin}, {Peng}, {Chou}, {Lu}, {Cheng}  \&
  {Ogilvie}}{{Lo} et~al.}{2015}]{Lo2015MNRAS.451..159L}
{Lo} J.-I.,  {Lin} M.-Y.,  {Peng} Y.-C.,  {Chou} S.-L.,  {Lu} H.-C.,  {Cheng}
  B.-M.,   {Ogilvie} J.~F.,  2015, \mn@doi [\mnras] {10.1093/mnras/stv935},
  \href {https://ui.adsabs.harvard.edu/abs/2015MNRAS.451..159L} {451, 159}

\bibitem[\protect\citeauthoryear{Lu, Chen  \& Cheng}{Lu
  et~al.}{2004}]{BingMing2004}
Lu H.-C.,  Chen H.-K.,   Cheng B.-M.,  2004, \mn@doi [Analytical chemistry]
  {10.1021/ac0494679}, 76, 5965

\bibitem[\protect\citeauthoryear{{Luna}, {Luna-Ferr{\'a}ndiz}, {Mill{\'a}n},
  {Domingo}, {Mu{\~n}oz Caro}, {Santonja}  \& {Satorre}}{{Luna}
  et~al.}{2017}]{Luna2017}
{Luna} R.,  {Luna-Ferr{\'a}ndiz} R.,  {Mill{\'a}n} C.,  {Domingo} M.,
  {Mu{\~n}oz Caro} G.~M.,  {Santonja} C.,   {Satorre} M.~{\'A}.,  2017, \mn@doi
  [\apj] {10.3847/1538-4357/aa7562}, \href
  {https://ui.adsabs.harvard.edu/abs/2017ApJ...842...51L} {842, 51}

\bibitem[\protect\citeauthoryear{{Mart{\'\i}n-Dom{\'e}nech},
  {Manzano-Santamar{\'\i}a}, {Mu{\~n}oz Caro}, {Cruz-D{\'\i}az}, {Chen},
  {Herrero}  \& {Tanarro}}{{Mart{\'\i}n-Dom{\'e}nech} et~al.}{2015}]{MD2015}
{Mart{\'\i}n-Dom{\'e}nech} R.,  {Manzano-Santamar{\'\i}a} J.,  {Mu{\~n}oz Caro}
  G.~M.,  {Cruz-D{\'\i}az} G.~A.,  {Chen} Y.~J.,  {Herrero} V.~J.,   {Tanarro}
  I.,  2015, \mn@doi [\aap] {10.1051/0004-6361/201526003}, \href
  {https://ui.adsabs.harvard.edu/#abs/2015A&A...584A..14M} {584, A14}

\bibitem[\protect\citeauthoryear{{Mart{\'\i}n-Dom{\'e}nech}, {Mu{\~n}oz Caro}
  \& {Cruz-D{\'\i}az}}{{Mart{\'\i}n-Dom{\'e}nech} et~al.}{2016}]{MD2016}
{Mart{\'\i}n-Dom{\'e}nech} R.,  {Mu{\~n}oz Caro} G.~M.,   {Cruz-D{\'\i}az}
  G.~A.,  2016, \mn@doi [\aap] {10.1051/0004-6361/201528025}, \href
  {https://ui.adsabs.harvard.edu/#abs/2016A&A...589A.107M} {589, A107}

\bibitem[\protect\citeauthoryear{{Mart{\'\i}n-Dom{\'e}nech}, {Cruz-D{\'\i}az}
  \& {Mu{\~n}oz Caro}}{{Mart{\'\i}n-Dom{\'e}nech} et~al.}{2018}]{MD2018}
{Mart{\'\i}n-Dom{\'e}nech} R.,  {Cruz-D{\'\i}az} G.~A.,   {Mu{\~n}oz Caro}
  G.~M.,  2018, \mn@doi [\mnras] {10.1093/mnras/stx2510}, \href
  {https://ui.adsabs.harvard.edu/\#abs/2018MNRAS.473.2575M} {473, 2575}

\bibitem[\protect\citeauthoryear{{McKay}, {Martin}, {Griffith}  \&
  {Keller}}{{McKay} et~al.}{1997}]{McKay1997}
{McKay} C.~P.,  {Martin} S.~C.,  {Griffith} C.~A.,   {Keller} R.~M.,  1997,
  \mn@doi [\icarus] {10.1006/icar.1997.5751}, \href
  {https://ui.adsabs.harvard.edu/\#abs/1997Icar..129..498M} {129, 498}

\bibitem[\protect\citeauthoryear{{Moore} \& {Hudson}}{{Moore} \&
  {Hudson}}{2003}]{Moore2003}
{Moore} M.~H.,  {Hudson} R.~L.,  2003, \mn@doi [\icarus]
  {10.1016/S0019-1035(02)00037-4}, \href
  {https://ui.adsabs.harvard.edu/\#abs/2003Icar..161..486M} {161, 486}

\bibitem[\protect\citeauthoryear{{Mu{\~n}oz Caro}, {Jim{\'e}nez-Escobar},
  {Mart{\'\i}n-Gago}, {Rogero}, {Atienza}, {Puertas}, {Sobrado}  \&
  {Torres-Redondo}}{{Mu{\~n}oz Caro} et~al.}{2010}]{Guille2010}
{Mu{\~n}oz Caro} G.~M.,  {Jim{\'e}nez-Escobar} A.,  {Mart{\'\i}n-Gago}
  J.~{\'A}.,  {Rogero} C.,  {Atienza} C.,  {Puertas} S.,  {Sobrado} J.~M.,
  {Torres-Redondo} J.,  2010, \mn@doi [\aap] {10.1051/0004-6361/200912462},
  \href {https://ui.adsabs.harvard.edu/#abs/2010A&A...522A.108M} {522, A108}

\bibitem[\protect\citeauthoryear{{Mumma} \& {Charnley}}{{Mumma} \&
  {Charnley}}{2011}]{Mumma2011}
{Mumma} M.~J.,  {Charnley} S.~B.,  2011, \mn@doi [Annual Review of Astronomy
  and Astrophysics] {10.1146/annurev-astro-081309-130811}, \href
  {https://ui.adsabs.harvard.edu/#abs/2011ARA&A..49..471M} {49, 471}

\bibitem[\protect\citeauthoryear{{{\"O}berg}, {Boogert}, {Pontoppidan},
  {Blake}, {Evans}, {Lahuis}  \& {van Dishoeck}}{{{\"O}berg}
  et~al.}{2008}]{Oberg2008}
{{\"O}berg} K.~I.,  {Boogert} A.~C.~A.,  {Pontoppidan} K.~M.,  {Blake} G.~A.,
  {Evans} N.~J.,  {Lahuis} F.,   {van Dishoeck} E.~F.,  2008, \mn@doi [\apj]
  {10.1086/533432}, \href
  {https://ui.adsabs.harvard.edu/\#abs/2008ApJ...678.1032O} {678, 1032}

\bibitem[\protect\citeauthoryear{{{\"O}berg}, {van Dishoeck}, {Linnartz}  \&
  {Andersson}}{{{\"O}berg} et~al.}{2010}]{Oberg2010}
{{\"O}berg} K.~I.,  {van Dishoeck} E.~F.,  {Linnartz} H.,   {Andersson} S.,
  2010, \mn@doi [\apj] {10.1088/0004-637X/718/2/832}, \href
  {https://ui.adsabs.harvard.edu/abs/2010ApJ...718..832O} {718, 832}

\bibitem[\protect\citeauthoryear{{{\"O}berg}, {Boogert}, {Pontoppidan}, {van
  den Broek}, {van Dishoeck}, {Bottinelli}, {Blake}  \& {Evans}}{{{\"O}berg}
  et~al.}{2011}]{Oberg2011}
{{\"O}berg} K.~I.,  {Boogert} A.~C.~A.,  {Pontoppidan} K.~M.,  {van den Broek}
  S.,  {van Dishoeck} E.~F.,  {Bottinelli} S.,  {Blake} G.~A.,   {Evans} II
  N.~J.,  2011, \mn@doi [\apj] {10.1088/0004-637X/740/2/109}, \href
  {https://ui.adsabs.harvard.edu/abs/2011ApJ...740..109O} {740, 109}

\bibitem[\protect\citeauthoryear{{Okabe}}{{Okabe}}{1978}]{Okabe1978}
{Okabe} H.,  1978, {Photochemistry of small molecules}

\bibitem[\protect\citeauthoryear{{Owen} et~al.,}{{Owen}
  et~al.}{1993}]{Owen1993}
{Owen} T.~C.,  et~al., 1993, \mn@doi [Science] {10.1126/science.261.5122.745},
  \href {https://ui.adsabs.harvard.edu/\#abs/1993Sci...261..745O} {261, 745}

\bibitem[\protect\citeauthoryear{{Shen}, {Greenberg}, {Schutte}  \& {van
  Dishoeck}}{{Shen} et~al.}{2004}]{Shen2004}
{Shen} C.~J.,  {Greenberg} J.~M.,  {Schutte} W.~A.,   {van Dishoeck} E.~F.,
  2004, \mn@doi [\aap] {10.1051/0004-6361:20031669}, \href
  {https://ui.adsabs.harvard.edu/#abs/2004A&A...415..203S} {415, 203}

\bibitem[\protect\citeauthoryear{Wong et~al.,}{Wong et~al.}{2017}]{Wong2017}
Wong M.~L.,  et~al., 2017, \mn@doi [Icarus]
  {https://doi.org/10.1016/j.icarus.2016.09.028}, 287, 110

\bibitem[\protect\citeauthoryear{{Wu}, {Wu}, {Chou}, {Lin}, {Lu}, {Lo}  \&
  {Cheng}}{{Wu} et~al.}{2012}]{Wu2012ApJ...746..175W}
{Wu} Y.-J.,  {Wu} C.~Y.~R.,  {Chou} S.-L.,  {Lin} M.-Y.,  {Lu} H.-C.,  {Lo}
  J.-I.,   {Cheng} B.-M.,  2012, \mn@doi [\apj] {10.1088/0004-637X/746/2/175},
  \href {https://ui.adsabs.harvard.edu/abs/2012ApJ...746..175W} {746, 175}

\bibitem[\protect\citeauthoryear{{Young} et~al.,}{{Young}
  et~al.}{2018}]{Young2018}
{Young} L.~A.,  et~al., 2018, \mn@doi [\icarus] {10.1016/j.icarus.2017.09.006},
  \href {https://ui.adsabs.harvard.edu/abs/2018Icar..300..174Y} {300, 174}

\bibitem[\protect\citeauthoryear{{d'Hendecourt}, {Allamandola}, {Grim}  \&
  {Greenberg}}{{d'Hendecourt} et~al.}{1986}]{d'Hendecourt1986A&A...158..119D}
{d'Hendecourt} L.~B.,  {Allamandola} L.~J.,  {Grim} R.~J.~A.,   {Greenberg}
  J.~M.,  1986, \aap, \href
  {https://ui.adsabs.harvard.edu/abs/1986A&A...158..119D} {158, 119}

\bibitem[\protect\citeauthoryear{{d'Hendecourt} et~al.,}{{d'Hendecourt}
  et~al.}{1996}]{d'hendecourt1996A&A...315L.365D}
{d'Hendecourt} L.,  et~al., 1996, \aap, \href
  {https://ui.adsabs.harvard.edu/abs/1996A&A...315L.365D} {315, L365}

\bibitem[\protect\citeauthoryear{{de Barros}, {Bordalo}, {Seperuelo Duarte},
  {da Silveira}, {Domaracka}, {Rothard}  \& {Boduch}}{{de Barros}
  et~al.}{2011}]{deBarros2011}
{de Barros} A.~L.~F.,  {Bordalo} V.,  {Seperuelo Duarte} E.,  {da Silveira}
  E.~F.,  {Domaracka} A.,  {Rothard} H.,   {Boduch} P.,  2011, \mn@doi [\aap]
  {10.1051/0004-6361/201016021}, \href
  {https://ui.adsabs.harvard.edu/\#abs/2011A&A...531A.160D} {531, A160}

\bibitem[\protect\citeauthoryear{{van Hemert}, {Takahashi}  \& {van
  Dishoeck}}{{van Hemert} et~al.}{2015}]{vanHemert2015}
{van Hemert} M.~C.,  {Takahashi} J.,   {van Dishoeck} E.~F.,  2015, \mn@doi
  [Journal of Physical Chemistry A] {10.1021/acs.jpca.5b02611}, \href
  {https://ui.adsabs.harvard.edu/abs/2015JPCA..119.6354V} {119, 6354}

\makeatother
\end{thebibliography}








\bsp	
\label{lastpage}
\end{document}